\documentclass[11pt]{article}
\def \d {{\rm d}}
\def \e {{\rm e}}

\begin{document}

\title{Null limits of generalised Bonnor--Swaminarayan solutions}

\author{J. Podolsk\'y\thanks{E--mail: {\tt Podolsky@mbox.troja.mff.cuni.cz}}
\\
\\ Institute of Theoretical Physics, Charles University,\\
V Hole\v{s}ovi\v{c}k\'ach 2, 18000 Prague 8, Czech Republic.\\
\\
and J. B. Griffiths\thanks{E--mail: {\tt J.B.Griffiths@Lboro.ac.uk}} \\ \\
Department of Mathematical Sciences, Loughborough University \\
Loughborough, Leics. LE11 3TU, U.K. \\ }

\maketitle

\begin{abstract}
\noindent 
The Bonnor--Swaminarayan solutions are boost-rotation symmetric space-times
which describe the motion of pairs of accelerating particles which are
possibly connected to strings (struts). In an explicit and unified form we
present a generalised class of such solutions with a few new observations.
We then investigate the possible limits in which the accelerations become
unbounded. The resulting space-times represent spherical impulsive
gravitational waves with snapping or expanding cosmic strings. We also
obtain an exact solution for a snapping string of finite length. 

\smallskip
\noindent {\bf Keywords}: Accelerating particles, cosmic strings,
impulsive spherical wave. 

\end{abstract}

%\pacs{04.20.Jb, 04.30.Nk, 11.10.Lm}

\section{Introduction}

We have recently shown \cite{PodGri99} that expanding impulsive
spherical gravitational waves may be considered as impulsive limits of
the Robinson--Trautman family of vacuum type~N space-times. Such waves
may be explicitly constructed using Penrose's method \cite{Pen72} of
cutting Minkowski (or de~Sitter or anti-de~Sitter) space along a null
cone and re-attaching the two parts with a suitable warp. However,
particular solutions with these properties may also be obtained as
limiting cases of certain known solutions which have boost-rotation
symmetry. (The first example of this procedure was presented by
Bi\v{c}\'ak and Schmidt at the end of section~5 of~\cite{BicSch89a}.) The
purpose of the present work is to list some particular boost-rotation
symmetric solutions in a suitable form, and to investigate their possible
impulsive wave limits in detail.

Boost-rotation symmetric space-times are the only explicitly known exact
solutions of Einstein's vacuum field equations which describing moving
particles, are radiative and asymptotically flat in the sense that they
admit global, though not complete, smooth null infinity, as well as smooth
spacelike and timelike infinities. The general properties of such
solutions have been extensively investigated and reviewed by Bi\v{c}\'ak
and Schmidt \cite{BicSch89b} (see also \cite{Bic85} and \cite{Bic87}), and
need not be repeated here.

Here, we will concentrate on the Bonnor--Swaminarayan (BS) solutions
\cite{BonSwa64} which are specific boost-rotation symmetric solutions.
These are particularly appropriate to consider in the null limit in which
impulsive waves will arise. However, some related solutions also have
similar limits. We will therefore review the BS and related solutions in a
unified way in the next two sections, adapting the notation to our
requirements and making a few new observations. We will then investigate
their impulsive limits in section~4.

\section{The Bonnor--Swaminarayan solutions}

The BS solutions \cite{BonSwa64} can be described by the line element
 \begin{equation}
 \d s^2=-\e^\lambda \d\rho^2 - \rho^2 \e^{-\mu} \d \phi^2 +
  (\zeta^2-\tau^2)^{-1} \left[\e^\mu (\zeta\d\tau-\tau\d\zeta)^2
-\e^\lambda(\zeta \d \zeta - \tau \d\tau)^2\right],
 \label{BSmetric}
 \end{equation}
 where
 \begin{eqnarray}
  \mu&=&-{2m_1\over A_1R_1} - {2m_2\over A_2R_2} +
   4m_1A_1 + 4m_2A_2 + B, \nonumber \\
  \lambda&=&8m_1m_2{A_1^3A_2^3(R_1-R_2)^2\over(A_2^2-A_1^2)^2R_1R_2}
  -{2m_1m_2\over A_1A_2R_1R_2}  \label{BSfunctions} \\
  &&-\left({m_1^2\over A_1^2R_1^4}
  + {m_2^2\over A_2^2R_2^4}\right)\rho^2(\zeta^2-\tau^2)
  + 2\left({m_1 A_1\over R_1}
  + {m_2 A_2\over R_2}\right)(\rho^2 + \zeta^2 - \tau^2) + B,  \nonumber
\end{eqnarray}
in which
\begin{equation}
 R_i={1\over 2}\sqrt{\left(\rho^2+\zeta^2-\tau^2-{1\over A_i^2}\right)^2
     + {4\over A_i^2}\rho^2} \qquad\qquad (i = 1, 2). \label{Ri}
\end{equation}
 For this line element, the metric functions $\mu$ and $\lambda$ depend
only on $\rho^2$ and $\zeta^2-\tau^2$. This dependence exhibits
explicitly the rotation and boost symmetry.

The detailed physical interpretation of this space-time was described in
\cite{BonSwa64} and \cite{Bon66} (a summary is given in
\cite{BoGrMa94}). In general, it represents the motion of two pairs of
uniformly accelerating particles which are possibly connected to conical
singularities (strings or struts). The particles are located
symmetrically, two on the positive and two on the negative $\zeta$-axis.
However, the point ``masses'' are not black holes, but are of the type
that are usually described as ``Curzon--Chazy'' particles. The
radiative properties of the BS solutions have been extensively studied by
Bi\v{c}\'ak \cite{Bic68,Bic71,Bic85,Bic87}.

\begin{figure}[hpt]      %BS spacetime diagram
\begin{center}
\setlength{\unitlength}{0.2mm}
\begin{picture}(200,200)(-100,-100)
\put(-100,0){\vector(1,0){200}}
\put(0,-100){\vector(0,1){200}}
\put(-100,-100){\line(1, 1){200}}
\put(-100, 100){\line(1,-1){200}}
\put(105,-4){$\zeta$}
\put(-4,105){$\tau$}
\thicklines
\qbezier(-110,100)(-10,0)(-110,-100)
\qbezier(-120,100)(-45,0)(-120,-100)
\qbezier(110,100)(10,0)(110,-100)
\qbezier(120,100)(45,0)(120,-100)
\end{picture}
\vskip-0.6cm
\end{center}

\caption{ A space-time diagram of the BS solutions}
\end{figure}
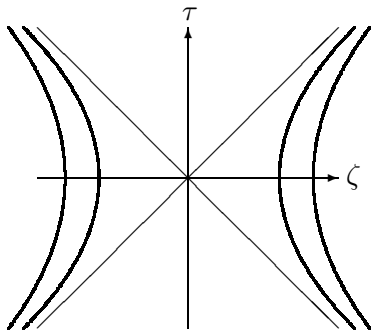

The above solution contains five arbitrary constants $m_1$, $m_2$, $A_1$,
$A_2$\  and \ $B$. These determine the mass and acceleration of each particle
and the singularity structure on the axis of symmetry $\rho=0$. Each pair
contains two particles with masses $m_i$. These are located on the axis at
the points where $R_i=0$. They thus have hyperbolic worldlines in the
$\tau$-$\zeta$ plane given by
 \begin{equation}
 \hbox{$m_i$ :}\qquad \rho=0, \quad \zeta=\pm\sqrt{\tau^2+{1\over
A_i^2}}\>.
 \label{positions}
 \end{equation}
 Each particle is uniformly accelerated with respect to a Minkowski
background, the accelerations being \ $\pm A_i$ \ in the $\pm\zeta$
directions  (we shall assume $A_2>A_1>0$). The minimum distance of each
particle from $\zeta=0$ is given by $\pm A_i^{-1}$.

The metric (\ref{BSmetric}) will generally contain conical singularities,
representing strings or struts, located on the axis of symmetry. These
will occur unless $\lambda+\mu\to0$ as $\rho\to0$ (see \cite{BicSch89b}). In
fact, it is possible to choose the constants $m_i$, $A_i$ and $B$ such that
this
regularity condition is satisfied on appropriate sections of the axis.

The following cases are of particular interest:

\bigskip\goodbreak\noindent{\bf Case 1}
\medskip

It is possible to make the choices
\begin{equation}
m_1={(A_2^2-A_1^2)^2\over 4A_1^3A_2^2}, \qquad
m_2=-{(A_2^2-A_1^2)^2\over4A_1^2A_2^3}, \qquad B=0.
\label{Case1}
\end{equation}
 With this, the axis is regular everywhere except at the locations of the
point particles.

%\begin{figure}[hpt]                  %Fig. for Case 1
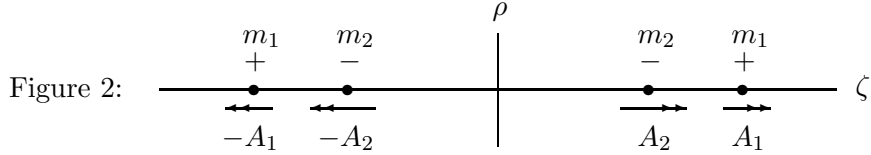
\begin{center}
\setlength{\unitlength}{0.25mm}
\begin{picture}(400,100)(0,-100)     %  {dim x, dim y}
\put(200,-20){\line(0,-1){60}}       % vertical axis of length 60
   \put(390,-53){$\zeta$}
\put(20,-50){\line(1,0){360}}        % horizontal axis of length 360
   \put(197,-10){$\rho$}
\put(70,-50){\circle*{6}}
   \put(64,-25){$m_1$}
   \put(65,-39){$+$}
\put(120,-50){\circle*{6}}
   \put(114,-25){$m_2$}
   \put(115,-39){$-$}
\put(280,-50){\circle*{6}}
   \put(274,-25){$m_2$}
   \put(275,-39){$-$}
\put(330,-50){\circle*{6}}
   \put(324,-25){$m_1$}
   \put(325,-39){$+$}
\put(80, -60){\vector(-1,0){25}}
\put(80, -60){\vector(-1,0){18}}
\put(135, -60){\vector(-1,0){35}}
\put(135, -60){\vector(-1,0){28}}
\put(265, -60){\vector(1,0){35}}
\put(265, -60){\vector(1,0){28}}
\put(320, -60){\vector(1,0){25}}
\put(320, -60){\vector(1,0){18}}
\put(48,-80){ $-A_1$}
\put(104,-80){$-A_2$}
\put(274,-80){$A_2$}
\put(324,-80){$A_1$}
\put(-60,-54){Figure 2:}
\end{picture}
\end{center}
\vskip-5mm
%\end{figure}

\noindent
This case had previously been described (in the static region) by Bondi
\cite{Bondi57} in 1957. For this choice of constants, the outer particles
have positive mass, and the inner particles negative mass. The
interactions between each pair of particles causes them both to accelerate
towards infinity.

\bigskip\goodbreak\noindent{\bf Case 2}
\medskip

The conditions for Case 1 can be relaxed by taking
\begin{equation}
m_1={(A_2^2-A_1^2)^2\over 4A_1^3A_2^2}, \qquad m_2 \ {\rm arbitrary},
\qquad B=0.
\end{equation}
 In this case, the axis is only regular between the particles of each
pair and between the two pairs of particles. However, the outer particles
are connected to infinity by two semi-infinite strings.

%\begin{figure}[hpt]               %Fig. for Case 2
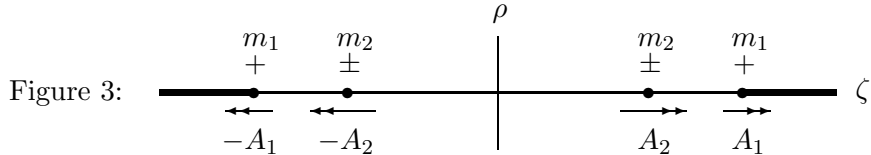
\begin{center}
\setlength{\unitlength}{0.25mm}
\begin{picture}(400,100)(0,-100)    %  {dim x, dim y}
\put(200,-20){\line(0,-1){60}}       % vertical axis of length 60
   \put(390,-53){$\zeta$}
\put(20,-50){\line(1,0){360}}        % horizontal axis of length 360
   \put(197,-10){$\rho$}
\put(20,-49.0){\line(1,0){50}}        % singularities
\put(20,-49.5){\line(1,0){50}}
\put(20,-50.5){\line(1,0){50}}
\put(20,-51.0){\line(1,0){50}}
\put(330,-49.0){\line(1,0){50}}        % singularities
\put(330,-49.5){\line(1,0){50}}
\put(330,-50.5){\line(1,0){50}}
\put(330,-51.0){\line(1,0){50}}
\put(70,-50){\circle*{6}}
   \put(64,-25){$m_1$}
   \put(65,-39){$+$}
\put(120,-50){\circle*{6}}
   \put(114,-25){$m_2$}
   \put(115,-39){$\pm$}
\put(280,-50){\circle*{6}}
   \put(274,-25){$m_2$}
   \put(275,-39){$\pm$}
\put(330,-50){\circle*{6}}
   \put(324,-25){$m_1$}
   \put(325,-39){$+$}
\put(80, -60){\vector(-1,0){25}}
\put(80, -60){\vector(-1,0){18}}
\put(135, -60){\vector(-1,0){35}}
\put(135, -60){\vector(-1,0){28}}
\put(265, -60){\vector(1,0){35}}
\put(265, -60){\vector(1,0){28}}
\put(320, -60){\vector(1,0){25}}
\put(320, -60){\vector(1,0){18}}
\put(48,-80){ $-A_1$}
\put(104,-80){$-A_2$}
\put(274,-80){$A_2$}
\put(324,-80){$A_1$}
\put(-60,-54){Figure 3:}
\end{picture}
\end{center}
\vskip-5mm
%\end{figure}

\noindent
In this case, the outer particles must have positive mass and are
``pulled'' towards infinity by the tension in the string. However, the
inner particles have arbitrary mass which may be positive, negative or
zero.

\bigskip\noindent{\bf Case 2a}
\medskip

Of particular interest is the special case of 2 in which $m_2=0$. For
this choice, the inner particles disappear, and the two remaining
particles are pulled towards infinity by semi-infinite strings. This
has been described by Bi\v{c}\'ak, Hoenselaers and Schmidt
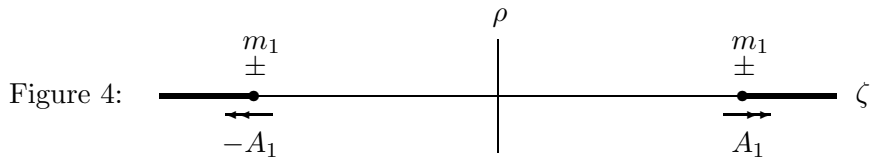
\cite{BicHoeSch83a}.

%\begin{figure}[hpt]               %Fig. for Case 2a
\begin{center}
\setlength{\unitlength}{0.25mm}
\begin{picture}(400,100)(0,-100)    %  {dim x, dim y}
\put(200,-20){\line(0,-1){60}}       % vertical axis of length 60
   \put(390,-53){$\zeta$}
\put(20,-50){\line(1,0){360}}        % horizontal axis of length 360
   \put(197,-10){$\rho$}
\put(20,-49.0){\line(1,0){50}}        % singularities
\put(20,-49.5){\line(1,0){50}}
\put(20,-50.5){\line(1,0){50}}
\put(20,-51.0){\line(1,0){50}}
\put(330,-49.0){\line(1,0){50}}        % singularities
\put(330,-49.5){\line(1,0){50}}
\put(330,-50.5){\line(1,0){50}}
\put(330,-51.0){\line(1,0){50}}
\put(70,-50){\circle*{6}}
   \put(64,-25){$m_1$}
   \put(65,-39){$\pm$}
\put(330,-50){\circle*{6}}
   \put(324,-25){$m_1$}
   \put(325,-39){$\pm$}
\put(80, -60){\vector(-1,0){25}}
\put(80, -60){\vector(-1,0){18}}
\put(320, -60){\vector(1,0){25}}
\put(320, -60){\vector(1,0){18}}
\put(48,-80){ $-A_1$}
\put(324,-80){$A_1$}
\put(-60,-54){Figure 4:}
\end{picture}
\end{center}
\vskip-5mm
%\end{figure}

\noindent
In this particular subcase, however, the restriction that the outer
particles must have positive mass no longer occurs. We only require
\begin{equation}
m_1 \ {\rm arbitrary}, \qquad m_2=0, \qquad B=0,
\label{case2a}
\end{equation}
 so that the remaining particles may also have negative mass, in which
case the conical singularity corresponds to a strut rather than a
string. (It may be noticed that this subcase can also be obtained in the
limit $A_2\to A_1$, in which case $m_1\to0$ and the masses at the end
of the conical singularity are replaced by the original inner masses
$m_2$ which may be arbitrary.)

\bigskip\goodbreak\noindent{\bf Case 3}
\medskip

A further case can be obtained by imposing the conditions that the axis
is regular only between each pair of particles and between each outer
particle and infinity. In this case there remains a string between the
two inner particles which must have negative mass. However, the
outer particles may have arbitrary (positive or negative) mass. The
parameters are given by
\begin{equation}
m_1 \ {\rm arbitrary}, \qquad m_2=-{(A_2^2-A_1^2)^2\over4A_1^2A_2^3}, \qquad
B=-4m_1A_1-4m_2A_2.
\label{case3}
\end{equation}

%\begin{figure}[hpt]               %Fig. for Case 3
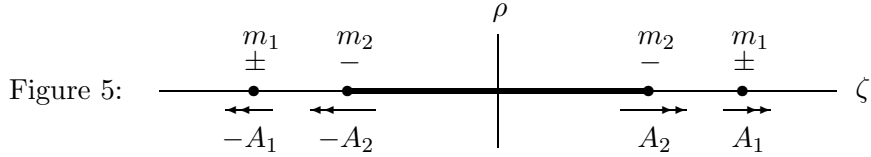
\begin{center}
\setlength{\unitlength}{0.25mm}
\begin{picture}(400,100)(0,-100)    %  {dim x, dim y}
\put(200,-20){\line(0,-1){60}}       % vertical axis of length 60
   \put(390,-53){$\zeta$}
\put(20,-50){\line(1,0){360}}        % horizontal axis of length 360
   \put(197,-10){$\rho$}
\put(120,-49.0){\line(1,0){160}}        % singularities
\put(120,-49.5){\line(1,0){160}}
\put(120,-50.5){\line(1,0){160}}
\put(120,-51.0){\line(1,0){160}}
\put(70,-50){\circle*{6}}
   \put(64,-25){$m_1$}
   \put(65,-39){$\pm$}
\put(120,-50){\circle*{6}}
   \put(114,-25){$m_2$}
   \put(115,-39){$-$}
\put(280,-50){\circle*{6}}
   \put(274,-25){$m_2$}
   \put(275,-39){$-$}
\put(330,-50){\circle*{6}}
   \put(324,-25){$m_1$}
   \put(325,-39){$\pm$}
\put(80, -60){\vector(-1,0){25}}
\put(80, -60){\vector(-1,0){18}}
\put(135, -60){\vector(-1,0){35}}
\put(135, -60){\vector(-1,0){28}}
\put(265, -60){\vector(1,0){35}}
\put(265, -60){\vector(1,0){28}}
\put(320, -60){\vector(1,0){25}}
\put(320, -60){\vector(1,0){18}}
\put(48,-80){ $-A_1$}
\put(104,-80){$-A_2$}
\put(274,-80){$A_2$}
\put(324,-80){$A_1$}
\put(-60,-54){Figure 5:}
\end{picture}
\end{center}
\vskip-5mm
%\end{figure}

\bigskip\goodbreak\noindent{\bf Case 3a}
\medskip

The special case of 3 arises if $m_1=0$. This
would result in two particles of negative mass connected by a string.
However, the restriction on $m_2$ no longer occurs and it may be chosen
arbitrarily (as in case 2a above). This has been described in
\cite{BicHoeSch83a}. Alternatively, we may obtain the same situation by
taking the limit as $A_2\to A_1$ in (\ref{case3}). This subcase is thus
given by
\begin{equation}
m_1 \ {\rm arbitrary}, \qquad m_2=0, \qquad B=-4m_1A_1.
\end{equation}

%\begin{figure}[hpt]               %Fig. for Case 3a
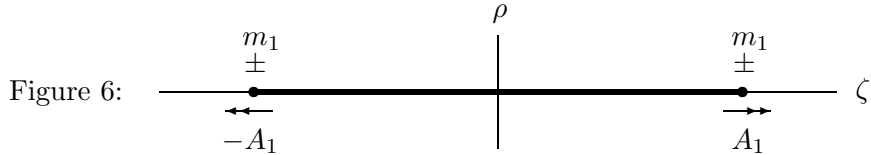
\begin{center}
\setlength{\unitlength}{0.25mm}
\begin{picture}(400,100)(0,-100)    %  {dim x, dim y}
\put(200,-20){\line(0,-1){60}}       % vertical axis of length 60
   \put(390,-53){$\zeta$}
\put(20,-50){\line(1,0){360}}        % horizontal axis of length 360
   \put(197,-10){$\rho$}
\put(70,-49.0){\line(1,0){260}}        % singularities
\put(70,-49.5){\line(1,0){260}}
\put(70,-50.5){\line(1,0){260}}
\put(70,-51.0){\line(1,0){260}}
\put(70,-50){\circle*{6}}
   \put(64,-25){$m_1$}
   \put(65,-39){$\pm$}
\put(330,-50){\circle*{6}}
   \put(324,-25){$m_1$}
   \put(325,-39){$\pm$}
\put(80, -60){\vector(-1,0){25}}
\put(80, -60){\vector(-1,0){18}}
\put(320, -60){\vector(1,0){25}}
\put(320, -60){\vector(1,0){18}}
\put(48,-80){ $-A_1$}
\put(324,-80){$A_1$}
\put(-60,-54){Figure 6:}
\end{picture}
\end{center}
\vskip-5mm
%\end{figure}

\medskip
Notice that for the special cases 2a and 3a describing only two
accelerated particles, the metric functions (\ref{BSfunctions}) can
be written in a simple form (see \cite{Bicak90})
 \begin{eqnarray}
  \mu&=&\mu_0=-{2m\over A R}+  4m A+ B, \nonumber \\
  \lambda&=&\lambda_0=-{m^2\over A^2R^4}\rho^2(\zeta^2-\tau^2)
  +{2m A\over R}(\rho^2 + \zeta^2 - \tau^2) + B,   \label{BSmono}
\end{eqnarray}
where $B=0$ for case 2a or $B=-4mA$ for case 3a,
\begin{equation}
 R={1\over 2}\sqrt{\left(\rho^2+\zeta^2-\tau^2-{1\over A^2}\right)^2
     + {4\over A^2}\rho^2}. \label{R}
\end{equation}
In these expressions, the unnecessary indices have been omitted
(i.e. $m=m_1,   A=A_1, R=R_1$).

\bigskip\goodbreak\noindent{\bf Case 4}
\medskip

A further case has been described in detail in \cite{BicHoeSch83b} for
which the axis is regular everywhere except between the two pairs of
particles. The parameters are here given by
 \begin{equation}
 m_1 \ {\rm arbitrary}, \qquad m_2=-{A_1\over A_2}m_1, \qquad B=0.
 \label{Case4}
 \end{equation}
In this case the masses of the two particles in each pair must be of
opposite sign. If the outer particle has positive (negative) mass, the
singularity corresponds to a strut (string).

%\begin{figure}[hpt]               %Fig. for Case 4
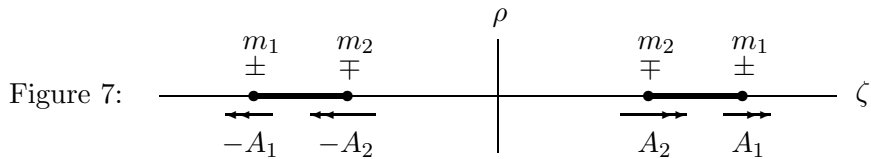
\begin{center}
\setlength{\unitlength}{0.25mm}
\begin{picture}(400,100)(0,-100)    %  {dim x, dim y}
\put(200,-20){\line(0,-1){60}}       % vertical axis of length 60
   \put(390,-53){$\zeta$}
\put(20,-50){\line(1,0){360}}        % horizontal axis of length 360
   \put(197,-10){$\rho$}
\put(70,-49.0){\line(1,0){50}}        % singularities
\put(70,-49.5){\line(1,0){50}}
\put(70,-50.5){\line(1,0){50}}
\put(70,-51.0){\line(1,0){50}}
\put(280,-49.0){\line(1,0){50}}        % singularities
\put(280,-49.5){\line(1,0){50}}
\put(280,-50.5){\line(1,0){50}}
\put(280,-51.0){\line(1,0){50}}
\put(70,-50){\circle*{6}}
   \put(64,-25){$m_1$}
   \put(65,-39){$\pm$}
\put(120,-50){\circle*{6}}
   \put(114,-25){$m_2$}
   \put(115,-39){$\mp$}
\put(280,-50){\circle*{6}}
   \put(274,-25){$m_2$}
   \put(275,-39){$\mp$}
\put(330,-50){\circle*{6}}
   \put(324,-25){$m_1$}
   \put(325,-39){$\pm$}
%\put(80,-32){\vector(0,1){24}}
%      \put(100,-200){\line(0,1){24}}
%
\put(80, -60){\vector(-1,0){25}}
\put(80, -60){\vector(-1,0){18}}
\put(135, -60){\vector(-1,0){35}}
\put(135, -60){\vector(-1,0){28}}
\put(265, -60){\vector(1,0){35}}
\put(265, -60){\vector(1,0){28}}
\put(320, -60){\vector(1,0){25}}
\put(320, -60){\vector(1,0){18}}
\put(48,-80){ $-A_1$}
\put(104,-80){$-A_2$}
\put(274,-80){$A_2$}
\put(324,-80){$A_1$}
\put(-60,-54){Figure 7:}
\end{picture}
\end{center}
\vskip-5mm
%\end{figure}

\section{The Bi\v{c}\'ak--Hoenselaers--Schmidt solutions}

Let us now consider some further solutions that can be obtained from
those above by taking suitable limits. A further solution will also be
included which contains the special cases above.

\bigskip\goodbreak\noindent{\bf Case 2b}
\medskip

One special limit (described in detail by Bi\v{c}\'ak, Hoenselaers and
Schmidt \cite{BicHoeSch83a}) of case 2 above that is of particular
interest is that in which $A_1\to0$. In this case, the outer particles,
and the strings attached to them, are scaled out to infinity. The
significance of this solution is that the remaining particles move freely
and are not connected to strings. However, they move under the action of
an exterior gravitational field for which the source is at infinity. It
may be noted that although $m_1\sim A_1^{-3}$ in this limit, $R_1\sim
A_1^{-2}$ and $\mu$ is bounded. The resulting solution is given by
 \begin{eqnarray}
 \mu&=&-{2m_2\over A_2 R_2}+  4m_2 A_2 +A_2^2\,(\rho^2-\zeta^2+\tau^2) ,
\label{Case2b}\\
\lambda&=&-\left({m_2^2\over A_2^2R_2^4}+A_2^4\right) \rho^2(\zeta^2-\tau^2)
 +A_2^2(\rho^2+\zeta^2-\tau^2) +{2m_2\over A_2R_2}(2A_2^2\rho^2+1) -4m_2A_2
\>. \nonumber
 \end{eqnarray}

%\begin{figure}[hpt]               %Fig. for Case 2b
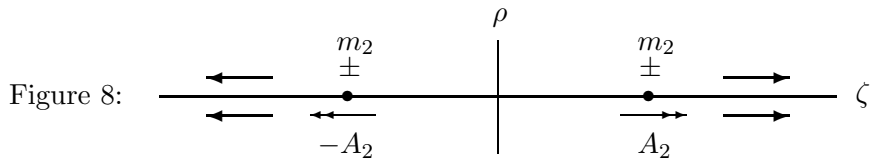
\begin{center}
\setlength{\unitlength}{0.25mm}
\begin{picture}(400,100)(0,-100)    %  {dim x, dim y}
\put(200,-20){\line(0,-1){60}}       % vertical axis of length 60
   \put(390,-53){$\zeta$}
\put(20,-50){\line(1,0){360}}        % horizontal axis of length 360
   \put(197,-10){$\rho$}
\put(120,-50){\circle*{6}}
   \put(114,-25){$m_2$}
   \put(115,-39){$\pm$}
\put(280,-50){\circle*{6}}
   \put(274,-25){$m_2$}
   \put(275,-39){$\pm$}
\put(135, -60){\vector(-1,0){35}}
\put(135, -60){\vector(-1,0){28}}
\put(265, -60){\vector(1,0){35}}
\put(265, -60){\vector(1,0){28}}
\put(104,-80){$-A_2$}
\put(274,-80){$A_2$}
\put(-60,-54){Figure 8:}
\thicklines
\put(80, -40){\vector(-1,0){35}}
\put(80, -60){\vector(-1,0){35}}
\put(320, -40){\vector(1,0){35}}
\put(320, -60){\vector(1,0){35}}
\end{picture}
\end{center}
\vskip-5mm
%\end{figure}

\bigskip\goodbreak\noindent{\bf Case 4a}
\medskip

We can also consider the limit of case 4 in which the $A_2\to A_1$ so
that the two particles in each pair approach each other. This procedure
has been described by Bi\v{c}\'ak, Hoenselaers and Schmidt
\cite{BicHoeSch83b}. It is convenient to introduce the parameter \ 
$M_{01}=2({m_1\over A_1}+{m_2\over A_2})$ \ which, with the constraint
(\ref{Case4}), becomes \ $M_{01}={2m_1\over A_1A_2^2}(A_2^2-A_1^2)$. \ In
this limit, Minkowski space-time is obtained unless the parameter $m_1$
is rescaled in such a way that $M_{01}$ is kept constant. This
particular solution is given by
\begin{eqnarray}
 \mu&=&-{M_{01}\over R}+{M_{01}\over 4A^2R^3}
  \left(\rho^2-\zeta^2+\tau^2+{1\over A^2}\right) , \nonumber \\
\lambda&=& -{M_{01}^2\over64R^8} \,\rho^2(\zeta^2-\tau^2)
\left[ \left((\rho^2+\zeta^2-\tau^2)^2-{1\over A^4}\right)^2
-{2\over A^4} \,\rho^2(\zeta^2-\tau^2)\right]
\label{Case4a} \\
&&\qquad -{M_{01}\over 4R^3}\left(\rho^2+\zeta^2-\tau^2\right)
  \left(\rho^2-\zeta^2+\tau^2+{1\over A^2}\right). \nonumber
 \end{eqnarray}
It may be noticed that an analogous limit of case~1 does not exist as
there is no freedom to rescale the mass $m_1$ (see (\ref{Case1})) and
Minkowski space is obtained as the two particles coalesce.

%\begin{figure}[hpt]               %Fig. for Case 4a
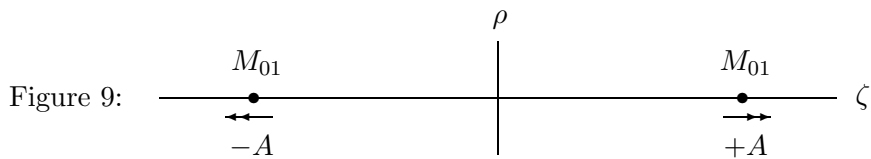
\begin{center}
\setlength{\unitlength}{0.25mm}
\begin{picture}(400,100)(0,-100)    %  {dim x, dim y}
\put(200,-20){\line(0,-1){60}}       % vertical axis of length 60
   \put(390,-53){$\zeta$}
\put(20,-50){\line(1,0){360}}        % horizontal axis of length 360
   \put(197,-10){$\rho$}
\put(70,-50){\circle*{6}}
   \put(58,-35){$M_{01}$}
\put(330,-50){\circle*{6}}
   \put(318,-35){$M_{01}$}
\put(80, -60){\vector(-1,0){25}}
\put(80, -60){\vector(-1,0){18}}
\put(320, -60){\vector(1,0){25}}
\put(320, -60){\vector(1,0){18}}
\put(52,-80){ $-A$}
\put(320,-80){$+A$}
\put(-60,-54){Figure 9:}
\end{picture}
\end{center}
\vskip-5mm
%\end{figure}

As shown in \cite{BicHoeSch83b}, the metric describes a combination of
monopole and dipole terms, so that the resulting particle has been
referred to as a Curzon--Chazy (01)-pole particle (for this reason we
have denoted the parameter by $M_{01}$). The space-time simply contains
two such particles. The situation looks similar to that for case 2b
above. However, in this case there is no external fields and the
particles are accelerated by their internal dipole component.

\bigskip\goodbreak\noindent{\bf Multipole particles}
\medskip

In \cite{BicHoeSch83b} other boost-rotationally symmetric spacetimes
generalising the Bonnor--Swaminarayan solution
(\ref{BSmetric}-\ref{BSfunctions}) were found. These represent the
fields of two accelerating particles with arbitrary multipole structure
attached to conical singularities as in the cases 2a and 3a above.
Moreover, for a special choice of parameters the space-times may be free
of conical singularities as in the case 4a, or 2b but without an
external field.

This class of solutions was presented in \cite{BicHoeSch83b} using prolate
spheroidal coordinates related to the Weyl form of the boost-rotational
symmetric metric. We will give here an explicit closed form of these
solutions in terms of the metric functions for the line element
(\ref{BSmetric}). However, the expression for $\lambda$ is presented here
in a much simpler form than that given in \cite{BicHoeSch83b}. The
derivation of this new expression, and the relation between the two forms,
are contained in the appendix. In addition, using the scaling property of
the Weyl metric, we can include the acceleration parameter~$A$ explicitly
(this parameter does not appear in \cite{BicHoeSch83b}, in which the
scaling was used to put $A={1\over2}$). In this new form, the general
class of multipole solutions is given by
\begin{eqnarray}
 \mu&=& 2\sum_{n=0}^{\infty}M_n \, {P_n\over{(x-y)^{n+1}} } + C,
\nonumber\\
\lambda&=&-2\sum_{k,l=0}^{\infty} M_k M_l \,{(k+1)(l+1)\over(k+l+2)}\,
   {(P_kP_l-P_{k+1}P_{l+1})\over {(x-y)^{k+l+2}}} \label{lambdamulti}\\
   &&\qquad\qquad -\left({x+y\over x-y}\right)
\sum_{n=0}^{\infty} {M_n\over 2^n}\,\sum_{l=0}^n
\left({2\over x-y}\right)^l P_l + D,
\nonumber
 \end{eqnarray}
where the constants $M_n$ represent the multipole moments,
the argument of the Legendre polynomials $P_n$ is $\alpha=(1-xy)/(x-y)$,
and $C$, $D$
are constants to be specified below. For the  prolate  spheroidal coordinates
$x$ and $y$ and for $\alpha$ in (\ref{lambdamulti}) one has to substitute from
the  relations
\begin{eqnarray}
&&x-y=4A^2R, \qquad\qquad x+y=2A^2(\rho^2+\zeta^2-\tau^2), \nonumber\\
&&\alpha={1\over 2R}\left(\rho^2-\zeta^2+\tau^2+{1\over A^2}\right),
 \label{alpha}
\end{eqnarray}
where $R$ is given by (\ref{R}). These imply the useful formulae
 $$ (x^2-1)(1-y^2)=16A^4\rho^2(\zeta^2-\tau^2)\label{x2-1}, \qquad
1+xy=2A^2(\zeta^2-\tau^2-\rho^2). $$

It now remains to set the values of the constants $C$ and $D$. To ensure
that the space-time is regular on the ``roof'' \ $\zeta^2-\tau^2=0$, \ it
is necessary that \ $\mu=\lambda$ \ at \ $\rho=0=\zeta^2-\tau^2$ \ (see
\cite{BicSch89b}). This yields explicitly that \
$D=C+\sum_{n=0}^{\infty}M_n/2^n$. \ Moreover, the metric is regular on the
axis provided \ $\mu+\lambda=0$ \ at \ $\rho=0$. \ Regularity of the axis
{\it between} the two particles, analogous to the case 2a (see Fig.~4),
requires that
\begin{equation}
 C=-\sum_{n=0}^{\infty} \ {M_n\over 2^n}, \qquad D=0. \label{stringout}
\end{equation}
 In this case there is generally a string connecting the particles to
infinity. The alternative situation, analogous to case 3a (see Fig.~6), in
which there is a string between the particles and the axis is regular
{\it outside} is given by
\begin{equation}
 C=0, \qquad D=\sum_{n=0}^{\infty} \ {M_n\over 2^n}. \label{stringin}
\end{equation}
The axis is obviously {\it regular everywhere}, except at the particle, if
the combination of multipole moments satisfies the condition that \
$\sum_{n=0}^{\infty}M_n/2^n=0$ \ so that \ $C=0=D$. \ This was pointed out
in \cite{BicHoeSch83b}.

Considering only the case $n=0$ in (\ref{lambdamulti}), we recover the
previous formula (\ref{BSmono}) for the {\it monopole} BS particles as in
case 2a or 3a with identification $M_0=-4mA$.

It is also straightforward to write an explicit solution representing
accelerated {\it dipole} particles with the single moment $M_1$:
 \begin{eqnarray}
  \mu_1&=&{M_1\over 16A^4R^3}\left(\rho^2-\zeta^2+\tau^2+{1\over A^2}\right)+
C, \nonumber \\
  \lambda_1&=&{M_1^2\over 512\,A^8R^4}\left(9\alpha^4-10\alpha^2+1\right)
\label{BSdip} \\
  &&\qquad- {M_1\over 4R}\left[1+{1\over4A^2R^2}
\left(\rho^2-\zeta^2+\tau^2+{1\over
A^2}\right)\right] \left(\rho^2+\zeta^2-\tau^2\right)
  + D, \nonumber
\end{eqnarray}

With these two observations, it may also be seen that case 4a given by
(\ref{Case4a}) is a special case of the general class of solutions
(\ref{lambdamulti}). It is a {\it combination} of both monopole and dipole
terms with the identification \ $2M_0=-M_1=-4A^2M_{01}$, \ so that the
constraint \ $M_0+M_1/2=0$ \ which guarantees the regularity of the axis is
automatically satisfied. This is consistent with the interpretation of the
solution (\ref{Case4a}) given in \cite{BicHoeSch83b}. Indeed, the metric
functions $\mu$ and $\lambda$ given by (\ref{Case4a}) can be written as \
$\mu=\mu_0+\mu_1$ \ and \ 
$\lambda=\lambda_0+\lambda_1+\lambda_{01}$, \ where $\mu_0, \lambda_0$ are
given by (\ref{BSmono}), $\mu_1, \lambda_1$ are  given by (\ref{BSdip}),
and \ $\lambda_{01}=M_{01}^2\alpha(1-\alpha^2)/4A^2R^3$.

\section{\bf Limiting cases of the above metrics}

\noindent{\bf Cases 3a and 2a}
\medskip

A particular limit of case 3a above in which $m_2=0$ and $A_1\to\infty$
was investigated by Bi\v{c}\'ak and Schmidt \cite{BicSch89a} in 1989. In
this limit, there are only two particles of vanishing mass and their
accelerations become infinite. In addition, it is necessary that the
parameter $m=m_1$ is scaled to zero in such a way that the ``monopole
moment'' $M_0=-4mA$ remains constant. The resulting space-time is flat
everywhere except on an expanding sphere and on the strut between the
particles. It therefore describes an expanding spherical impulsive
gravitational wave that is generated by the two particles which move apart
at the speed of light in a Minkowski background and are connected to each
other by an expanding strut.

An analogous situation in which (for $\tau>0$) two null particles recede
from the origin with the speed of light and are connected to infinity by
semi-infinite strings was constructed explicitly by Gleiser and Pullin
\cite{GlePul89} in 1989. This describes the space-time representing an
expanding impulsive spherical gravitational wave generated by a snapping
cosmic string. Such a solution can also be obtained as a limit of case 2a
above \cite{Bicak90}. Again, it is necessary that $m$ is scaled to zero such
that $M_0$ remains constant. However, as pointed out by Bi\v{c}\'ak
\cite{Bicak90}, the complete solution rather describes two semi-infinite
strings approaching at the speed of light and separating again at the
instant at which they collide.

In the remainder of this section, we will investigate similar null limits
of all the remaining cases described in sections 2 and 3.

\bigskip\goodbreak\noindent{\bf Multipole particles}
\medskip

We may first consider the null limit of the general class of solutions
(\ref{lambdamulti}-\ref{alpha}). These represent two accelerated particles
with {\it arbitrary multipole structure} possibly with strings between the
particles or with strings connecting each particle to infinity. Here, we
let $A\to\infty$ while all the multipole moments $M_n$ are kept {\it
constant}. In this limit we obtain
\begin{eqnarray}
&& 2R\to |\,\rho^2+\zeta^2-\tau^2\,|,\quad\qquad\qquad
\alpha\to{{\rho^2-\zeta^2+\tau^2}\over
{\left|\,\rho^2+\zeta^2-\tau^2\,\right|}}
, \nonumber\\
&&x-y \to 2A^2\,|\,\rho^2+\zeta^2-\tau^2\,|, \qquad
  x+y \to 2A^2\,(\,\rho^2+\zeta^2-\tau^2\,), \label{limits} \\
&&  {x+y \over x-y} \to {\rm sign}\,(\,\rho^2+\zeta^2-\tau^2\,). \nonumber
\end{eqnarray}
 In this limit, $\mu$ in (\ref{lambdamulti}) approaches the constant $C$.
Also the first term for $\lambda$ vanishes, and only the contribution $l=0$
in the second term remains finite. Consequently, the metric functions for
the solutions which we obtain in the above null limit can be written as
\begin{eqnarray}
\mu&=& C\> , \nonumber\\
\lambda&=& D-{\rm sign}\,(\,\rho^2+\zeta^2-\tau^2\,)\,
  \sum_{n=0}^{\infty} \ {M_n\over 2^n}\>. \label{mulambda}
\end{eqnarray}

For the particular values of the constants $C$ and $D$ given by
(\ref{stringout}), this solution appears to be a generalisation of the null
limit of case 2a which describes a situation of a snapping cosmic string.
The ends of two semi-infinite cosmic string move in opposite directions
with the speed of light,  generating an impulsive spherical gravitational
wave. However, in this limit, {\it the multipole structure of the initial
particles disappears} and the solution is characterised by the single
constant \ $M=\sum_{n=0}^{\infty}\ {M_n/ 2^n}$. \ Thus, the null limit for
any accelerating particle with a multipole structure is identical to that
for a particle with just a monopole term $M_0=M$.

Considering the alternative values of the constants given by
(\ref{stringin}), we obtain a null limit in which a spherical impulsive
wave is generated by a finite string whose length is expanding at the
speed of light. However, at the ends of the strings the arbitrary
multipole structure of the initial ``particles'' described by the
multipole moments $M_n$ again disappears. This null limit is thus identical
to that of the case 3a above as originally obtained in \cite{BicSch89a}.

Performing the well-known transformation (see e.g. \cite{BicSch89b})
 \begin{equation}
 \rho=\textstyle{1\over2}(v-u), \qquad
\tau =\pm\textstyle{1\over2}(v+u)\,\cosh\chi, \qquad
\zeta=\textstyle{1\over2}(v+u)\,\sinh\chi, 
 \end{equation}
 we can put the solution (\ref{mulambda}) into the standard form of
boost-rotational symmetric metrics with null coordinates $u$ and $v$,
\begin{equation}
 \d s^2=\e^\lambda \,\d u\,\d v - \textstyle{1\over4}(v-u)^2 \,\e^{-\mu}
\,\d\phi^2
 - \textstyle{1\over4}(v+u)^2 \,\e^\mu \,\d \chi^2 \,.
 \label{BSnull}
\end{equation}
 Since \ $\rho^2+\zeta^2-\tau^2=-uv$, \ the metric functions take the
following form:
\begin{eqnarray}
&& {\rm snapping\ string:\ \>}\qquad\mu= -\sum_{n=0}^{\infty} \ {M_n\over
2^n} \> , \qquad
\lambda=\left[\,\Theta(uv)-\Theta(-uv)\,\right]\>\sum_{n=0}^{\infty} \
{M_n\over 2^n} \>,  \nonumber\\
&& {\rm expanding\ string:}\qquad\mu= 0\> , \qquad\qquad\qquad
\lambda=2\,\Theta(uv)\>\sum_{n=0}^{\infty} \ {M_n\over 2^n} \>, \nonumber
\end{eqnarray}
where $\Theta$ is the Heaviside step function. Notice that in both cases
$\mu$ is a constant, but there is a discontinuity in the otherwise
constant value of $\lambda$ with the step \ 
$2\,\sum_{n=0}^{\infty}\>{M_n/2^n}$ \ on the null cone \ $uv=0$. \ 
However, it is possible to find a transformation to coordinates in
which the metric is {\it continuous} everywhere. In the region where the
functions $\mu$ and $\lambda$ are  constant, the transformation
\begin{equation}
{\cal U}= {\textstyle{1\over2}}\,u\,\e^{\lambda+\mu/2}, \qquad
{\cal V}= {\textstyle{1\over2}}\,v\,\e^{-\mu/2}, \qquad
\psi= \chi\,\e^\mu,
\label{trans}
\end{equation}
brings the line element (\ref{BSnull}) into the form
\begin{equation}
 \d s^2=4 \,\d {\cal U} \,\d {\cal V} - {\cal A}^2 \,\d\phi^2 
- {\cal B}^2 \,\d\psi^2 ,
 \label{GP}
\end{equation}
where
\begin{equation}
 {\cal A}={\cal V}-{\cal U}\>\e^{-(\lambda+\mu)}\>,\qquad
 {\cal B}={\cal V}+{\cal U}\>\e^{-(\lambda+\mu)}\>.
 \label{AB}
\end{equation}
This metric is explicitly continuous, including on the null cone given by
\ ${\cal U}=0$, \ even if there has been a discontinuity in $\lambda$ at \ 
$uv=0$ \ in the original coordinates. Such a step is removed by the
compensating discontinuity in the expression for ${\cal U}$ in
(\ref{trans}).

The solution for a {\it snapping string} can thus be written in the
continuous form (\ref{GP}). By substituting the corresponding $\lambda$
and $\mu$ into (\ref{AB}), and restricting to the region $v>0$, we obtain
\begin{eqnarray}
 {\cal A}&=&{\cal V}-\left[\,\Theta({\cal U})+\Theta(-{\cal U})
    \exp\left(2\sum_{n=0}^{\infty} \ {M_n\over 2^n}\right)\right]{\cal
U}\>,\nonumber\\
 {\cal B}&=&{\cal V}+\left[\,\Theta({\cal U})+\Theta(-{\cal U})
    \exp\left(2\sum_{n=0}^{\infty} \ {M_n\over 2^n}\right)\right]{\cal U}\>.
 \label{ABsnap}
\end{eqnarray}
 The metric (\ref{GP}), (\ref{ABsnap}) is exactly of the type constructed
previously by Gleiser and Pullin \cite{GlePul89} by a different method. It
represents an impulsive spherical gravitational wave  propagating in the
Minkowski universe. However, outside the wave (where ${\cal U}<0$), there
extends the ``snapped" cosmic string which is characterized by a deficit
angle $(1-\beta)2\pi$, where here \
$\beta=\exp(\sum_{n=0}^{\infty}\>{M_n/2^n})$. \ This generalizes the
previous result presented in \cite{Bicak90} which was obtained as the null
limit of case 2a describing two accelerated monopole particles. And, as in
this particular case, the complete solution has to be extended
symmetrically to negative times, as has been argued in
\cite{Bicak90}.

Similarly, we can write the solution for an {\it expanding string} in the
form (\ref{GP}):
\begin{eqnarray}
 {\cal A}&=&{\cal V}-\left[\,\Theta({\cal
U})\exp\left(-2\sum_{n=0}^{\infty} \ {M_n\over 2^n}\right)+\Theta(-{\cal
U})\right]{\cal U}\>,\nonumber\\
 {\cal B}&=&{\cal V}+\left[\,\Theta({\cal
U})\exp\left(-2\sum_{n=0}^{\infty} \ {M_n\over 2^n}\right)+\Theta(-{\cal
U})\right]{\cal U}\>.
 \label{ABexp}
\end{eqnarray}
In this case there is a string with the deficit angle $(1-\beta^{-1})2\pi$
in the Minkowski space inside the  impulsive wave in the region
${\cal U}>0$.

There exists an alternative continuous form of the above
solutions. By applying the transformation (\ref{trans}) followed by \
$U=-2\,{\cal U}$, \ $V={\cal V}$, \ $Z={1\over\sqrt2}(\psi+{\rm i}\,\phi)$,
\ we may convert the metric (\ref{BSnull}) with constant $\lambda$ and
$\mu$, in the region $u<0$, $v>0$, into the form
\begin{equation}
 \d s^2=-2\,|\,V \d Z + U H \d \bar Z\,|^2 - 2\,\d  U \,\d  V\>,
 \label{NutPena}
\end{equation}
where \ $H=-{1\over2}\,\e^{-(\lambda+\mu)}$ \ and \ $U>0$. \ Performing now
a different transformation
\begin{eqnarray}
 U &=& -\> {uv\over u+v}\,\exp
  \left[ {\textstyle{\lambda\over2}} -\chi\,e^{(\mu-\lambda)/2} \
\right]
\>,  \nonumber\\
 V &=& {\textstyle{1\over2}}\,(u+v)\,\exp
  \left[ {\textstyle{\lambda\over2}}+\chi\,e^{(\mu-\lambda)/2} \ \right]
\>,  \label{trans2b}\\
 Z &=& {\textstyle{1\over\sqrt2}}\,{v-u\over v+u}\,\exp\left[
  -\chi\,e^{(\mu-\lambda)/2}
  + {\rm i}\,\phi\,e^{-(\mu+\lambda)/2} \ \right] \>,  \nonumber
\end{eqnarray}
of the metric (\ref{BSnull}) in the region  $u>0$, $v>0$, where $U<0$, we
obtain
\begin{equation}
 \d s^2=-2\,V^2 \,\d Z \,\d \bar Z - 2\,\d  U \,\d  V\>.
 \label{NutPenb}
\end{equation}
It is now obvious that we can match the two metrics (\ref{NutPena}) and
(\ref{NutPenb}) across the null cone $u=0$, which corresponds to $U=0$,
so that the resulting metric
\begin{equation}
 \d s^2=-2\,|\,V\,\d Z +U\Theta(U)\,H\,\d\bar Z\,|^2 -2\,\d U\,\d V\>,
 \label{NutPen}
\end{equation}
is continuous. This metric is contained within a general class of metrics
which describe impulsive spherical gravitational waves (see \cite{Pen72},
\cite{NutPen92}, \cite{Hogan93} and \cite{PodGri99}). In the general case,
the function $H(Z)$ can be obtained as the Schwarzian derivative of a
``warp'' function $h(Z)$ which permits a geometrical interpretation
\cite{PodGri00}. The above solution for a snapping string is given
by \ $h=\e^{\beta z}$, \ where \
\hbox{$\beta=\sqrt2\,\exp(\sum_{n=0}^{\infty}{M_n/2^n})$} \ corresponding
to \ $H=-{1\over2}\exp(2\sum_{n=0}^{\infty}{M_n/2^n})$. The alternative
case for an expanding string is given by \ $\beta\to\beta^{-1}$ \
corresponding to \
\hbox{$H=-{1\over2}\exp(-2\sum_{n=0}^{\infty}{M_n/2^n})$}.

\bigskip\goodbreak\noindent{\bf An alternative limit and Case 4a}
\medskip

It can immediately be observed that if the condition \
$\sum_{n=0}^{\infty}\>{M_n/2^n}=0$ \ is satisfied (i.e. when the original
metric is regular everywhere on the axis except at the locations of the
particles), the corresponding null limit results in a trivial Minkowski
spacetime without an impulsive wave. However, it is also possible in this
case to perform a {\it more involved null limit} which results in a
different class of spacetimes than those described by (\ref{mulambda}).
The sum in the second term for $\lambda$ in (\ref{lambdamulti}) can be
written as
 \begin{equation}
\sum_{n=0}^{\infty} {M_n\over 2^n}\,\sum_{l=0}^n
\left({2\over x-y}\right)^l P_l
 = \sum_{n=0}^{\infty} {M_n\over 2^n}
 +\left({2\alpha\over x-y}\right) \sum_{n=1}^{\infty} {M_n\over 2^n}
 +\sum_{l=2}^n\left({2\over x-y}\right)^l P_l
\sum_{n=l}^{\infty} {M_n\over 2^n}.
 \label{expand}
 \end{equation}
 When $\sum_{n=0}^{\infty}\>{M_n/2^n}=0$, the second term in the expansion
(\ref{expand})
 \begin{equation}
\left({2\alpha\over x-y}\right) \sum_{n=1}^{\infty} {M_n\over 2^n}
 =-{2M_0\,\alpha\over{x-y}}
 \end{equation}
 becomes dominant, and a different null limit can be obtained. Considering
that \ $x-y\sim A^2$ \ in the null limit $A\to\infty$,  we may here rescale
the remaining monopole moment $M_0$ in such a way that \ 
$M=M_0 A^{-2}$ \ remains constant. With these assumptions and using
(\ref{limits}) we can write the final form of the non-trivial null limit as
\begin{eqnarray}
 \mu&=&{M\over \left|\,\rho^2+\zeta^2-\tau^2 \,\right|} , \nonumber \\
\lambda&=& -M^2{\rho^2(\zeta^2-\tau^2) \over (\,\rho^2+\zeta^2-\tau^2\,)^4}
  + M{\rho^2-\zeta^2+\tau^2 \over (\,\rho^2+\zeta^2-\tau^2\,)^2}\>
  {\rm sign}\,(\,\rho^2+\zeta^2-\tau^2\,)\>. \label{2ndlimit}
 \end{eqnarray}
 This represents a unique alternative solution which can be obtained in the
null limit described above if \ $\sum_{n=0}^{\infty}\>{M_n/2^n}=0$, \ after
a suitable rescaling of the multipole moments. However, even in this more
involved limit, the  multipole  structure is again completely ``erased''.
Moreover, the metric is not acceptable as a description of an impulsive
gravitational wave since it is singular on the null cone \ 
$\rho^2+\zeta^2=\tau^2$. \ It may further be noted that when \ $M_0=0$, \
the solution is simply Minkowski space.

Notice finally, that the solution (\ref{2ndlimit}) exactly corresponds to that
which is obtained in the null limit $A\to\infty$ from the metric
(\ref{Case4a}) of case 4a with the identification $M=-2M_{01}$.

\bigskip\goodbreak\noindent{\bf Cases 2b, 1, 2 and 3}
\medskip

Concerning case 2b, it may immediately be seen from (\ref{Case2b}) that the
null limit in which $A_2\to\infty$ does not exist. The terms in the metric
functions which represent the external field responsible for the
acceleration diverge.

We have now considered all the above special cases which involve just two
accelerating particles. We therefore return to the remaining cases which
involve two pairs of distinct particles. In all these cases, it is only
possible to consider null limits as $A_2\to\infty$ (since $A_2>A_1$). In
cases 1 and 2, this limit diverges ($\mu\sim A_2^2$). However, in case~3,
the limit for $\mu$ is given by
 \begin{eqnarray}
  \mu&=&-{2m_1\over A_1R_1} - {1\over A_1^2|\rho^2+\zeta^2-\tau^2|},
\nonumber
\end{eqnarray}
 and this can be seen to diverge on the expanding spherical surface
$\rho^2+\zeta^2=\tau^2$. Null limits of the cases 1, 2 and 3 must
therefore be considered to be unphysical.

\bigskip\goodbreak\noindent{\bf Case 4}
\medskip

It now only remains to consider the limit of case~4. We again consider the
limit as $A_2\to\infty$, but we can now do this while scaling $m_2$ to
zero in such a way that $A_2m_2=-A_1m_1=$ constant. In this case the limit
is given by
 \begin{eqnarray}
  \mu&=&-{2m\over A R} \>, \nonumber \\
  \lambda&=&-{m^2\over A^2R^4}\rho^2(\zeta^2-\tau^2)
  +{2m A\over R}(\rho^2 + \zeta^2 - \tau^2)
 -{\rm sign}(\rho^2 + \zeta^2 - \tau^2)\,4mA,
\label{step}
\end{eqnarray}
 in which $R=R_1$ is given by (\ref{R}), and the subscript 1 has been
omitted from $m_1$, $A_1$ and $R_1$. This solution contains two free
parameters $m$ and $A$ (although it is not possible to consider the limit
as $A\to0$ while keeping $m$ finite). It represents a snapping string of
{\it finite length} whose outer ends are accelerating apart as shown in
figure~10.
\medskip

%\begin{figure}[hpt]      %BS spacetime diagram
\begin{center}
\setlength{\unitlength}{0.2mm}
\begin{picture}(200,200)(-100,-100)
\put(-100,0){\vector(1,0){200}}
\put(0,-100){\vector(0,1){200}}
\put(-100,-100){\line(1, 1){200}}
\put(-100, 100){\line(1,-1){200}}
\put(105,-4){$\zeta$}
\put(-4,105){$\tau$}
\thicklines
\qbezier(-120,100)(-45,0)(-120,-100)
\qbezier(120,100)(45,0)(120,-100)

\put( 0, 0){\line(1, 0){83}}
\put(10,10){\line(1, 0){74}}
\put(20,20){\line(1, 0){65}}
\put(30,30){\line(1, 0){56}}
\put(40,40){\line(1, 0){49}}
\put(50,50){\line(1, 0){42}}
\put(60,60){\line(1, 0){37}}
\put(70,70){\line(1, 0){31}}
\put(80,80){\line(1, 0){27}}
\put(90,90){\line(1, 0){23}}

\put(  0, 0){\line(-1, 0){82}}
\put(-10,10){\line(-1, 0){72}}
\put(-20,20){\line(-1, 0){63}}
\put(-30,30){\line(-1, 0){56}}
\put(-40,40){\line(-1, 0){48}}
\put(-50,50){\line(-1, 0){42}}
\put(-60,60){\line(-1, 0){35}}
\put(-70,70){\line(-1, 0){31}}
\put(-80,80){\line(-1, 0){26}}
\put(-90,90){\line(-1, 0){23}}

\put(10,-10){\line(1, 0){74}}
\put(20,-20){\line(1, 0){65}}
\put(30,-30){\line(1, 0){56}}
\put(40,-40){\line(1, 0){49}}
\put(50,-50){\line(1, 0){42}}
\put(60,-60){\line(1, 0){37}}
\put(70,-70){\line(1, 0){31}}
\put(80,-80){\line(1, 0){27}}
\put(90,-90){\line(1, 0){23}}

\put(-10,-10){\line(-1, 0){72}}
\put(-20,-20){\line(-1, 0){63}}
\put(-30,-30){\line(-1, 0){56}}
\put(-40,-40){\line(-1, 0){49}}
\put(-50,-50){\line(-1, 0){42}}
\put(-60,-60){\line(-1, 0){35}}
\put(-70,-70){\line(-1, 0){31}}
\put(-80,-80){\line(-1, 0){26}}
\put(-90,-90){\line(-1, 0){23}}

\end{picture}
\vskip-0.3cm
\end{center}

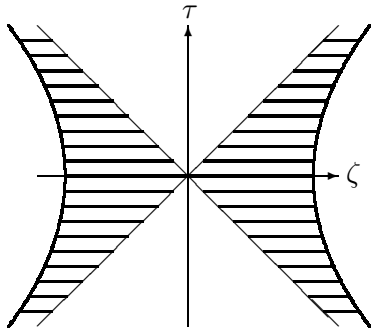
\centerline{Figure 10. A space-time diagram for a finite snapping string}
\bigskip

The two outer particles of mass $m$ are caused to accelerate by the strut
or string (there is a deficit angle if $m<0$) connecting them. However, in
this case, the finite strut breaks (or the string snaps) at its midpoint
and the two broken ends separate at the speed of light.

Apart from the step change in $\lambda$, the metric (\ref{step}) is
identical to that of cases 2a and 3a as given in (\ref{BSmono}). For the
general family of metrics (\ref{BSmetric}), the curvature tensor components
are linear in $\lambda$, and its discontinuities give rise to impulsive
components on the null cone. In fact, the Ricci tensor components still
vanish everywhere. However, an impulsive component arises in the Weyl
tensor on the null cone\footnote{These statements apply also to the null
multipole limits (\ref{mulambda}) as well as to the null limits of cases
2a and 3a.}. This is interpreted as describing an expanding spherical
impulsive gravitational wave that is generated by the snapping of the
string.

The complete solution of course is time symmetric as illustrated in
figure~10. This situation resembles that of the known solutions for
snapping infinite cosmic strings \cite{GlePul89}, \cite{Bicak90}. However,
this solution differs in that it is not flat on both sides of the
impulsive wave.

\section{Conclusion}

We have reviewed the Bonnor--Swaninarayan family of boost-rotationally
symmetric solutions which was generalised by Bi\v{c}\'ak, Hoenselaers and
Schmidt. We have written these solutions in a unified way paying particular
attention to certain limiting cases. These solutions describe the
accelerated motion of pairs of particles possibly attached to conical
singularities on the axis of symmetry. For the case of multipole
particles, we have presented a new simpler form of the solution.

We have specifically investigated the possible null limits of these
solutions. In most cases, the limits are trivial or not physically
acceptable. However, in some cases we have obtained physically interesting
limits which describe snapping or expanding cosmic strings generating
spherical impulsive gravitational waves. For multipole particles, we have
shown that the internal multipole structure vanishes in this limit, leaving
the same solution for a snapping or expanding cosmic string. We have also
presented a new solution for a snapping cosmic string of finite length.

\section*{Acknowledgements}

We are grateful to Professor J. Bi\v{c}\'ak for suggesting that we
investigate this topic and for his helpful comments on the draft of the
paper. This work was supported by a visiting fellowship from the Royal
Society and, in part, by the grant GACR-202/99/0261 of the Czech Republic.

\section*{Appendix}

The solution (\ref{lambdamulti}) for multipole particles was first
presented in \cite{BicHoeSch83b} using prolate spheroidal coordinates. The
expression for $\mu$ and the first term for $\lambda$ in
(\ref{lambdamulti}) are known from the standard Weyl solutions. However,
the second term in $\lambda$ arises as a `mixed' term representing the
interaction between the multipole terms and the `boost potential'.
Denoting the interaction term for the $n^{\rm th}$ multipole by
$\lambda_n$, this is required to satisfy the equation
 \begin{equation}
\partial_\beta\,\lambda_n =-{2(n+1)\over\sqrt{\beta^2-4\alpha\beta+4}}
\left({P_{n+1}\over\beta^{n+1}}-{2P_n\over\beta^{n+2}}\right),
 \label{eqn3.11}
 \end{equation}
 (this is a correction of equation (3.11) in \cite{BicHoeSch83b} in which
${1\over2}\lambda_n=\gamma_{{\rm m},n}-\beta^{-(n+1)}P_n$)
 where the independent coordinates are \ $\alpha=(1-xy)/(x-y)$ \ and \
$\beta=x-y$, \ and the Legendre polynomials have argument $\alpha$. We
have found that this can be integrated directly by introducing the
function 
 \begin{equation}
 \Lambda_n=\lambda_n-\textstyle{1\over2}\lambda_{n-1}.
 \label{Lambdan}
 \end{equation}
 Using the recurrence relation for Legendre polynomials and 
(\ref{eqn3.11}), we find that $\Lambda_n$ must satisfy 
 $$ \partial_\beta\,\Lambda_n = -\partial_\beta \left(
\beta^{-(n+1)} \sqrt{\beta^2-4\alpha\beta+4} \ P_n \right), $$ 
 which can immediately be integrated. Setting the integration constant to
zero and using a recurrence formula derived from (\ref{Lambdan}) with
$\lambda_0=-\beta^{-1}\sqrt{\beta^2-4\alpha\beta+4}$, we finally
obtain the integral 
 $$ \lambda_n= -{1\over 2^n} \>
\beta^{-1}\sqrt{\beta^2-4\alpha\beta+4} \>
 \sum_{l=0}^n \left({2\over\beta}\right)^l P_l\>. $$ 
 In prolate spheroidal coordinates, this is 
 \begin{equation}
  \lambda_n= -{1\over 2^n} \left({x+y\over x-y}\right)
 \sum_{l=0}^n \left({2\over x-y}\right)^l P_l\>,
 \label{gamma1}
 \end{equation}
 which, with the coefficient $M_n$, gives the second term in
(\ref{lambdamulti}) above.

The equivalent expression presented in \cite{BicHoeSch83b}, equation 
(3.13), is 
 \begin{equation}
  \lambda_n=  - {1\over 2^n}\,\left({x+y\over x-y}\right)
\,(n+1) (P_nf_{n+1}-P_{n+1}f_n),
\label{gamma2}
 \end{equation}
where
$$f_{n+1}=\sum_{k=0}^n{1\over k+1} P_{n-k} \sum_{l=0}^k
\left({2\over x-y}\right)^l P_{k-l}.$$
Note that in equation (3.13) of \cite{BicHoeSch83b}, there is also a
term \ $P_nf_{n+1}(\infty)-P_{n+1}f_n(\infty)$, \ 
where \ $f_{n+1}(\infty)=\sum_{k=0}^n{1\over k+1} P_{n-k} P_k$. \ However,
using the identity \ $(n+1)(P_{n+1}Q_n-P_nQ_{n+1})=1$ \ and the standard
definition \
$Q_n(\alpha)={1\over2}P_n(\alpha)\ln[(1+\alpha)/(1-\alpha)]-f_n(\infty)$,
\ it can easily be shown that \
$(n+1)[P_nf_{n+1}(\infty)-P_{n+1}f_n(\infty)]=1$. \ This term thus
represents only an additive constant which can be omitted.

Since both the solutions (\ref{gamma1}) and (\ref{gamma2}) satisfy the
same differential  equations, they can differ only by a constant. However,
by considering the particular  value at $x=1$, $y=0$, it can be seen that
this constant must be zero. {}From this we may now deduce the relation
 $$ (n+1)(P_nf_{n+1}-P_{n+1}f_n)=\sum_{l=0}^n \left({2\over x-y}\right)^l
P_l \>. $$ 
   By expanding the left-hand side in powers of \ $2/(x-y)$ \ and
comparing coefficients, we obtain the following non-trivial identity
for the Legendre polynomials.

\bigskip\noindent{\bf Theorem:}
For non-negative integers $m$ and $n$, with $m\le n$, define the
functions $X_n^{(m)}(\alpha)$ by
 $$ X_n^{(m)}\equiv\sum_{k=m}^n \textstyle{1\over k+1}\> P_{n-k} P_{k-m}
\ , $$ 
\indent\indent\indent
where $P_n(\alpha)$ are Legendre polynomials, and $X_{m-1}^{(m)}=0$. Then
 \begin{equation}
 \textstyle{ P_nX_n^{(m)}-P_{n+1}X_{n-1}^{(m)}={1\over n+1}\>P_m} 
 \end{equation}
\indent\indent\indent
for arbitrary $m$ and $n$ and for arbitrary value of the argument $\alpha$.

\bigskip\noindent
The left-hand side of the above identity is a polynomial of order $2n-m$.
It can therefore be expanded as a series of Legendre polynomials. However,
remarkably, all the coefficients vanish identically except for one.

\end{document}